# GRAVITY: beam stabilization and light injection subsystems

O. Pfuhl*[a], M. Haug[a], F. Eisenhauer[a], D. Penka[a], A. Amorim[b], S. Kellner[a], S. Gillessen[a], T. Ott[a], E. Wieprecht[a], E. Sturm[a], F. Haußmann[a], M. Lippa[a]

[a]Max-Planck-Institut für extraterrestrische Physik, Giessenbachstraße, 85748 Garching, Germany;
[b]SIM, Fac. de Ciências da Univ. de Lisboa, Campo Grande, Edif. C1, P-1749-016 Lisbon, Portugal;

## ABSTRACT

We present design results of the 2nd generation VLTI instrument GRAVITY beam stabilization and light injection subsystems. Designed to deliver micro-arcsecond astrometry, GRAVITY requires an unprecedented stability of the VLTI optical train. To meet the astrometric requirements, we have developed a dedicated 'laser guiding system', correcting the longitudinal and lateral pupil position as well as the image jitter. The actuators for the correction are provided by four 'fiber coupler' units located in the GRAVITY cryostat. Each fiber coupler picks the light of one telescope and stabilizes the beam. Furthermore each unit provides field de-rotation, polarization analysis as well as atmospheric piston correction. Using a novel roof prism design offers the possibility of on-axis as well as off-axis fringe tracking without changing the optical path. Finally the stabilized beam is injected with minimized losses into single-mode fibers via parabolic mirrors. We present lab results of the first guiding- as well as the first fiber coupler prototype regarding the closed loop performance and the optical quality. Based on the lab results we discuss the on-sky performance of the system and the implications concerning the sensitivity of GRAVITY.

*pfuhl@mpe.mpg.de

## 1. INTRODUCTION

The second generation instrument GRAVITY for the Very Large Telescope Interferometer (VLTI) will open a new era of high-resolution imaging and narrow-angle precision astrometry. It will be able to combine four telescopes, the 1.8m Auxiliary Telescopes (AT) or the 8m Unit Telescopes (UT) of the ESO observatory at Paranal. The large collecting area of the UTs and the capability of fringe-tracking will allow phase-referenced imaging of sources as faint as $m_K$ ~16 at a 3 milli-arcsecond resolution as well as micro-arcsecond precision astrometry. The improvement in angular resolution will be about a factor 15 compared to diffraction-limited imaging on current 10m class telescopes. In its dual-field mode, GRAVITY will provide relative position measurements between two objects with an accuracy of 10μas, i.e. an angle equal to a Euro coin diameter seen at the distance of the moon. Measuring at that accuracy level means observing the universe in motion, since 10 μas/yr correspond to 5m/s at a distance of 100pc (close star-forming regions) and 50 km/s at a distance of 1Mpc (Andromeda galaxy). GRAVITY has been proposed in 2005 as an adaptive optics assisted beam combiner for the second generation VLTI instrumentation[1]. The instrument will provide high precision narrow-angle astrometry and phase-referenced interferometric imaging at wavelengths between 1.95μm and 2.45μm. A bright wavefront reference star is picked with the PRIMA star separator and imaged onto the GRAVITY infrared wavefront sensor located in the VLT Coud´e laboratory. Alternatively stars, bright in the visible bands, can be used in combination with the optical wavefront sensor MACAO. The wavefront correction of either sensor is applied to the deformable mirror of the UT and provides a diffraction-limited image. The 2" FoV of the VLTI are re-imaged via the main delay lines in the GRAVITY beam combiner instrument. Laser guiding beams are launched at the telescope spiders and at the star separators to track pupil and tip/tilt perturbations in the VLTI tunnel. Sensors and actuators contained in the beam combiner instrument sense the motions and apply the corresponding corrections. Slow image drifts on sky are meanwhile monitored by the acquisition camera[4]. This camera also measures pupil motion lateral and longitudinal with respect to the optical axis. The stabilized beam is then separated by the fiber coupler into the science- and the fringe-tracking object and coupled individually into single-mode fibers. The polarization and the differential optical path between the two objects is then adjusted by the fiber control unit using fibered differential delay lines (FDDL) and polarization rotators. The actual beam superposition happens in an integrated optics (IO) chip[2], one for each object. Imprinted on the chip is a four phase-shift fringe sampling, that provides instantaneous phase and visibility information for all baselines. The bright reference object feeds the fringe-tracking spectrometer[3]. Here the phase and group delay is computed from five spectral channels at a rate of few hundred Hz. The corresponding OPD correction is applied to a piezo actuator contained in the fiber coupler. This allows stabilizing the fringes of the faint object and to use integration times significantly longer than the atmospheric coherence time. The science spectrometer[3], optimized for long, background-limited observations of the faint object offers a variety of observing modes. This includes low (R ~ 22) up to intermediate spectral resolution (R ~ 4500) modes. Additionally, both spectrometers offer the possibility to split and analyze linear polarization. The internal differential OPD of the two objects is measured with a dedicated laser metrology[5]. The laser is injected at the level of the IO chip and follows the optical path of the objects in reverse direction. The metrology interference pattern is then sampled after the primary mirror of the telescopes. The metrology signal is probed with a common phase-shifting technique in combination with photo diodes mounted to the telescope spiders. GRAVITY provides simultaneously differential phase and visibility information for at least five spectral channels and two objects. This allows interferometric imaging and astrometry on six baselines, using the four Auxiliary- or the four Unit Telescopes.

## 2. FIBER COUPLER SUBSYSTEM

### 2.1 Overview

The instrument GRAVITY will contain four fiber coupler units. The main purpose of the units is to pick up and stabilize the beams from the four telescopes, to split the FoV and to feed the light of two objects into single-mode fibers. Each fiber coupler contains the optics, motors and piezo actuators to de-rotate the field and to stabilize the beam tip/tilt and pupil wander. A rotatable half-wave plate allows adjusting and analyzing linear polarization. An internal cold stop reduces the thermal background. For calibration purposes, each unit is equipped with a retro-reflector that allows mapping the GRAVITY fibers onto the acquisition camera. The internal optical path between the integrated optics and the fiber coupler is monitored by a metrology reference diode, providing the OPD feedback for the fibered delay lines. Each unit allows separating two objects (science and fringe tracking object) within the 2" FoV (UT). In this dual-field mode, the two objects are coupled into two single-mode fibers. After passing the differential delay lines and the polarization control unit, the light of each object coming from the four telescopes interferes in the integrated optics chip. The bright object is used to correct the atmospheric piston jitter and to stabilize the fringes of the second (faint) object. This allows integrating significantly longer than the atmospheric coherence time on the faint object. Instead of a few milli-seconds, the integration time can be as long as minutes, increasing the sensitivity by orders of magnitude. Additionally, the fiber coupler offers a single-field mode, where fringe-tracking and science integration is done on the same object. This particular mode can be chosen for bright objects, where high resolution spectroscopy is desired. The mechanical structure of the fiber coupler is mounted on the optical bench in the cryostat and kept at a constant temperature of 240K. The temperature is a compromise between the thermal background and the operability of the stepper motors and piezo actuators. The latter are required for the fast OPD actuation (fringe-tracking) and to control tip/tilt and pupil wander of the incident beam. The tip/tilt stabilization, operated in closed loop, is driven by the acquisition camera and the guiding system. The guiding system tracks fast perturbations in the tunnel using laser beacons, while the acquisition camera guides slowly on a reference star. Another piezo actuator allows controlling the pupil in the instrument. Driven by the acquisition camera, this actuator corrects pupil wander and alignment. The joint field and pupil stabilization is required to ensure an optimum coupling efficiency and to reduce the tip/tilt and pupil induced astrometric error.

### 2.2 Optical layout

The fiber coupler is designed to feed the beam from one telescope into the single-mode fibers of the beam combiner instrument. Based on the optical interface of the VLTI[6], the design has to accept an 18mm beam and a 0.247° FoV (corresponds to 2" /UT on sky). Given that the AT FoV angle in the lab is always smaller than the UT angle, in the following we always refer to the UT case, if not stated otherwise. Since the vignetted FoV of the VLTI is somewhat larger, we assumed for the design an oversized FoV of 8" on sky. Imaging the extended FoV on the acquisition camera eases the identification of objects in crowded fields and relaxes the requirements on the telescope pointing accuracy. Furthermore, the clear aperture of all surfaces is matched to a pupil, oversized by ~20% to be able to track and correct lateral pupil motion. The operating wavelengths spans from the visible to the IR. For the various control-loops, we use two lasers at 658nm and 1200nm, and for acquisition and science we use the astronomical H- and K-band. Consequently, the fiber coupler throughput is optimized at these wavelengths.

Each fiber coupler unit contains the following optical and mechanical elements:

- K-mirror mounted on a rotation stage, to de-rotate the FoV
- Half-wave plate, optimized for operation in K-band (1.95-2.45μm) and at a temperature of -33°C
- Piezo Z/tip/tilt platform, to correct piston and tip/tilt perturbations
- Piezo tip/tilt platform, to control lateral pupil motion
- Dichroic, splitting the guiding- and acquisition wavelength from the science wavelength (reflect $1.90 < \lambda < 2.45$ μm; transmit $1.50 < \lambda < 1.80$ & 1.20 & 0.65μm)
- Retro-reflector, to map the fiber position on the acquisition camera
- Metrology diode, to monitor the internal optical path between the fiber coupler and the IO
- Roof-prism to split the field (dual-field mode) or to act as a beam splitter (single-field mode)
- Off-axis parabolic mirrors, made from diamond-turned aluminum, to focus and collimate the beam

The bulk of the fiber coupler consists of aluminum. The complete structure, as well as the off-axis parabolic mirrors is made from aluminum. This avoids differential thermal contraction, keeping the optical properties constant and the system always in focus. The aluminum mirrors have an additional nickel coating that allows accurate diamond turning of the mirror surface. On top of the nickel, a layer of gold ensures high reflectivity in the infrared. Using off-axis parabolas avoids lenses and the corresponding chromatic errors in the design. The layout of the fiber coupler is shown in Figure 1. Although formally not part of the fiber coupler, the entrance window of the cryostat is included in the figure. After passing the $CaF_2$ window, the beam passes the K-mirror. The assembly of three flat mirrors is mounted on a rotation stage. It serves as a derotator that compensates the field rotation induced by the telescope motion. After the K-mirror, a half-wave plate (HWP) is introduced in the optical path. Also mounted on a rotation stage, the HWP allows rotating the linear polarization of the incoming light. In combination with a wollaston prism in the spectrometer, this allows analyzing the object polarization. The entrance pupil of the instrument is placed on a flat mirror that is actively controlled by a piezo actuator. This actuator provides the fast tip/tilt and piston control (TTP) and it serves as a field selector. The second piezo actuator is located in an intermediate field plane. It allows moving the pupil within the instrument. Following a parabolic mirror, a dichroic splits the collimated beam into the science band and the guiding and acquisition band. The H-band and optical light is transmitted to the acquisition camera and guiding system receiver. The science wavelength (K-band) is reflected to the fiber injection optics. The fiber injection optics consists of a camera that re-images the FoV onto the roof-prism. Here the field is split into two objects. Using another two parabolic mirrors each object is re-imaged and de-magnified at the position of the fibers. The fibers are mounted on XYZ translation stages to focus and center the objects on the fibers. Behind the dichroic, a combination of an aspheric lens and a flat mirror acts as a retro-reflector for the metrology light. The dichroic is tuned such that ~99% of the light is reflected to the telescope but about 1% passes the dichroic, is retro-reflected and propagated to the acquisition camera. This allows re-imaging the metrology light from the fibers onto the camera. Knowing the fiber position on the camera allows a precise positioning of the object image on the fiber.

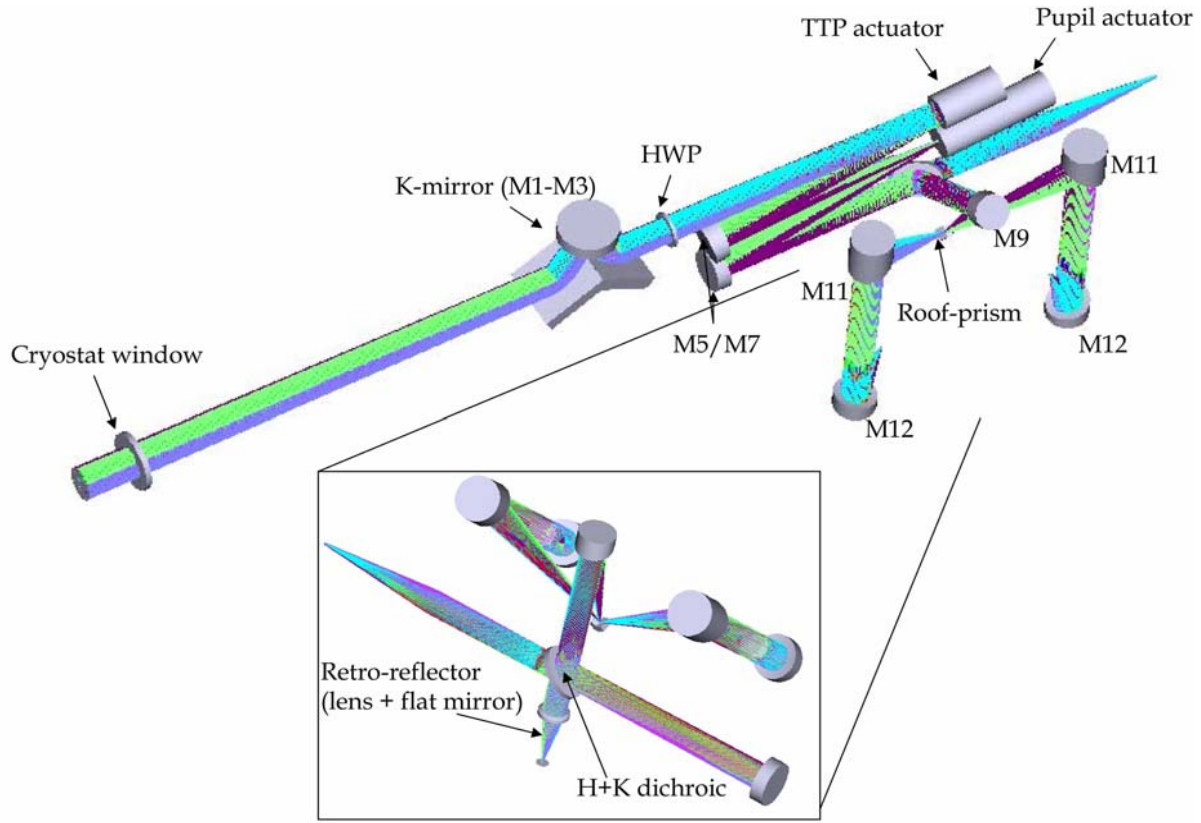

Figure 1. Optical design of the fiber coupler. The beam coming from one telescope enters the cryostat through a CaF2 window. The rotatable K-mirror compensates the image rotation due to the telescope orientation. The half-wave plate (HWP) controls the polarization orientation. The tip/tilt-piston (TTP) and pupil actuators stabilize the incident beam. A dichroic in the collimated beam splits the acquisition (H-band) and guiding light from the science light (K-band). The science light is reflected to the roof-prism, where the field is split in science and fringe-tracking object. Each object is coupled via a parabolic relay into individual single-mode fibers. The acquisition and guiding (laser) light passes the dichroic and enters the acquisition camera. In order to align the objects with the fibers, the retro-reflector allows mapping the fiber position on the acquisition camera. The metrology reference diode monitors the optical path between the IO and the fiber coupler.

## 2.3 Dual-field versus single-field mode

Among the main requirements for the fiber coupler is to provide a dual-field as well as a single-field mode. In order to avoid optic wheels or other movable parts, the goal was to design an optical element such that the two modes can be offered without additional actuators. The field splitting (dual-field) is done by introducing a roof-prism, i.e. two angled flat mirrors with a sharp edge into a field plane. Starting from that basic concept, a special roof-prism was designed that offers both field- and beam-splitting capability. The prism substrate is fused silica, transparent in the IR. The roof surface of the prism consists of coated quadrants. Two gold coated quadrants constitute the typical roof-prism, i.e. two angled mirrors. The tilted mirrors split the field in two half's. The other two quadrants are coated with a beam-splitter- and an anti-reflective (AR) coating. The bottom of the prism is again coated with gold. Thus in the single-field mode, the beam is split 50/50 at the first surface. The transmitted part is reflected at the bottom and exits the prism through the AR coated surface. The angles of the prism are tuned such that the exiting beam travels the same path as the dual-field mode. The beam traveling through the prism is slightly defocused due to the extended path. However, this can be partially compensated by re-focusing the fiber mounted on the translation stage. The roof-prism concept is illustrated in Figure 2. The f-number at the roof prism is F# = 5.6 corresponding to a diffraction limited point spread function (PSF) diameter of ~15μm in the K-band. The sharpness of the roof prism edge, determines the minimum object separation. It is limited by manufacturing processes and < 10μm, therefore smaller than the diffraction limit. Since the roof-prism is located in a field plane and the quadrants are only spaced by an equivalent of 1" on sky (0.16mm), changing the mode only requires offsetting the internal tip/tilt actuator. Figure 3 shows a picture of the 10 delivered roof-prism and a zoom-in under the microscope.

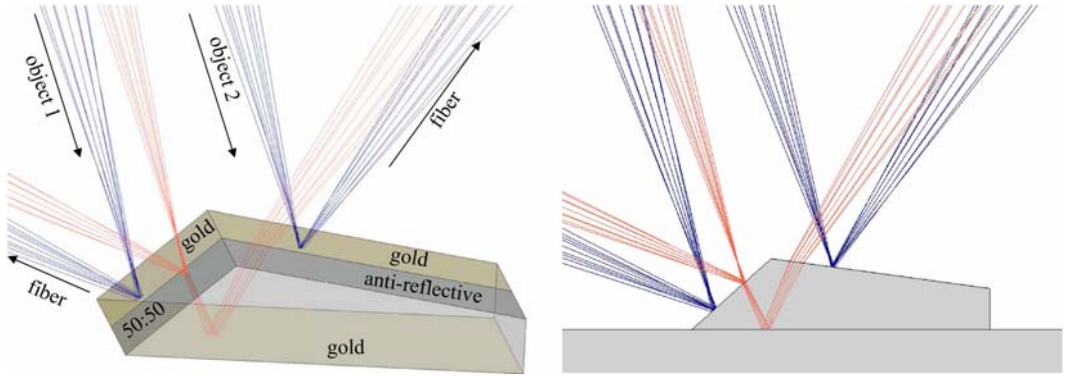

Figure 2. Left: 3D drawing of the roof-prism and the two operation modes (single/dualfield). In the dual-field mode two gold-coated quadrants split the field and reflect the light of two objects (blue rays) in opposite directions. The single-field mode splits the the light of one object (red rays) at the first surface. It reflects 50% of the light and transmits the other 50%. The transmitted light is reflected at the bottom of the roof-prism and exits the glass substrate at the opposite side of the prism. The following optical path is identical for both modes. Via a parabolic mirror relay, the two objects (or the split object) are coupled into two single-mode fibers. Right: roof-prism seen from the side.

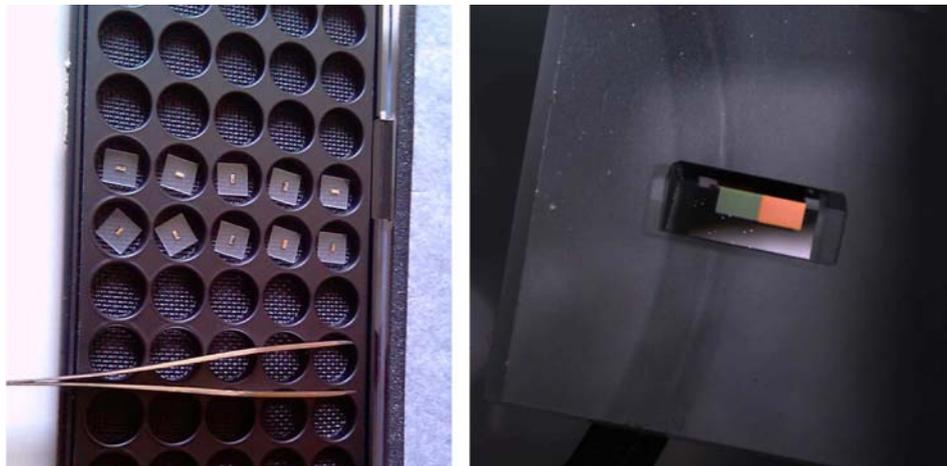

Figure 3. Picture of the 10 roof-prisms delivered from Zuend Precision Optics. Each prism is glued on a $10 \times 10mm^2$ N-BK7 glass plate for handling purposes. Right: One prism under the microscope. The prism base is $4 \times 1.5mm^2$. The optical active area however is the central $0.3 \times 0.3mm^2$.

**2.4 Optical performance**

**2.4.1 Image quality**

In the fiber coupler we abandoned lenses in favor of off-axis parabolic mirrors. Using mirrors avoids typical lens aberrations such as chromaticity. Typical drawbacks of parabolas are off-axis aberrations. The prime condition for any parabola is that beams parallel to the axis of symmetry are focused to a common point, independent of beam diameter. However, beams not parallel to the axis suffer from coma. The aberration gets worse, the larger the off-axis angle is which substantially limits the FoV with acceptable wavefront error. Yet, it is possible to compensate aberrations introduced by one parabolic mirror with a second one by tuning the focal lengths and off-axis distances of the relay. The first parabolic relay consisting of the mirrors M5 and M7 (see Figure 1) is such an example. Both mirrors have the same focal length and off-axis distances, yet with opposite sign. This makes the relay symmetric with respect to a normal vector through the intermediate field center, i.e. the symmetry axis of the pupil actuator. This symmetry cancels any aberrations apart from surface defects, providing a perfect collimated beam after M7. The combination of the three parabolas M9, M11 and M12 is more difficult and is governed by three conditions. The parabola M9 focuses and splits the beam at the roof-prism. The off-axis distance of M9 is determined by the minimum space required to fit the roof-prism next to the dichroic mount. After the roof-prism, the two emerging beams are collimated by two identical M11 mirrors. The focal length of the last parabola M12 is fixed to f = 104.35mm in order to match the optimum f-number F# = 2.5 (determined by the mode-field radius of the fibers, $\omega_B$=3.83µm @ K-band). The off-axis distance of M12 is determined by the minimum space required to mount a fiber at the focus position. Given these preconditions, the only free parameter is the off-axis distance of M11. Since the off-axis directions of M9 and M12 are perpendicular to each other, the off-axis position of M11 also has two be adjusted in X and Y direction. Using the wavefront error as a minimization criterion, the Zemax optimization routine was used to find the optimum off-axis distance. The resulting optical performance is excellent for a large FoV. Figure 4 shows the calculated wavefront error as function of field angle. The nominal dual-field wavefront error is close to zero at the field center and the maximum wavefront error for all wavelengths across the ±2" UT FoV on sky (±0.247° lab) is < 0.025 waves, corresponding to a Strehl ratio > 98%. Across 1" the Strehl is even > 99%. In case of the single-field mode the optical quality of the science and the fringe-tracking channel is different. While the reflected beam has the same optical quality as in the dual-field case, the beam propagating through the prism suffers from defocus and chromatic aberration. At 1" the Strehl already drops to 91%. Yet, in the single-field mode, only one object is needed in the FoV. This allows operating close to the optical axis with a wavefront error of the order 0.023 waves (Strehl > 98%). In that sense, the single-field mode performance is practically not degraded compared to the dual-field mode.

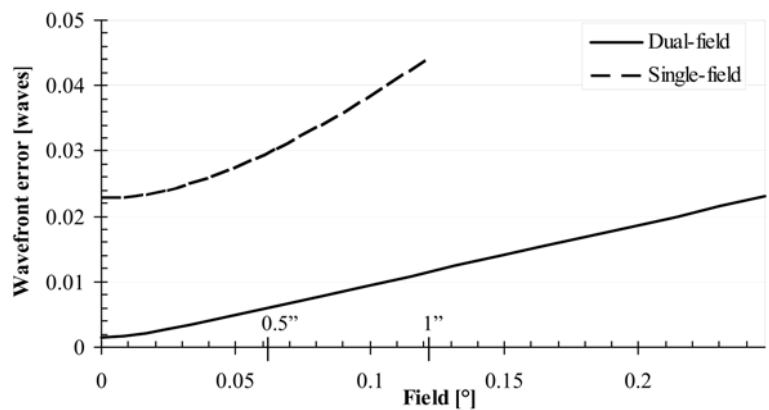

Figure 4. Nominal wavefront error of the dual-field mode as function of field angle in the 18mm beam (solid line). Two tickmarks show the equivalent 0.5" and 1" field on sky (UT). The dashed line shows the wavefront error for the beam propagated through the roof-prism in the single-field mode. The reflected beam has the same error as in the dual-field case.

**2.4.2 Transmission**

Any stellar interferometer is a complex optical system that requires delay lines and numerous other optical functions. Therefore optical throughput is always a major concern in these systems. For that reason, we designed the fiber coupler with regard to high throughput in the science band (1.9−2.45µm). The bulk absorption in MgF2, fused silica, and quartz is negligible in K-band (see e.g. product information of Crystran Ltd.; UK). Therefore only reflection losses have to be taken into account. Each fiber coupler unit comprises five glass surfaces, front- and back-surface of the cryostat window, the same for the half-wave plate and the tip of the single-mode fiber. For all glass surfaces inside vacuum, we assumed anti-reflective coatings with a typical loss of 0.5% per surface (according to various suppliers). Since the front surface of the cryostat window is exposed to ambient, we assumed in this case an increased loss due to dust of 1.5%. With the exception of the dichroic (with a loss of 1%), all other surfaces are gold-coated mirrors; in total ten reflections with a reflectivity of > 99% each in K-band. Multiplying the individual component losses yields a throughput of 85.9%. Taking into account the toleranced Strehl of 94%, the nominal coupling efficiency of 77.8% and some additional margin of 95%, the total throughput is ~60%.

## 2.5 Opto-mechanics

### 2.5.1 Structure

Each of the four fiber coupler units is self-contained. The overall size of one unit is 230 × 604×372mm$^3$. Each box consists of more than 700 individual pieces and weighs 25 kg. The complete structure and several of the optical elements are manufactured from aluminum (AlMg4.5Mn0.7). The choice of the material matches the cryostat optical bench to ensure a temperature independent behavior, when cooled down to the operating temperature of 240K.

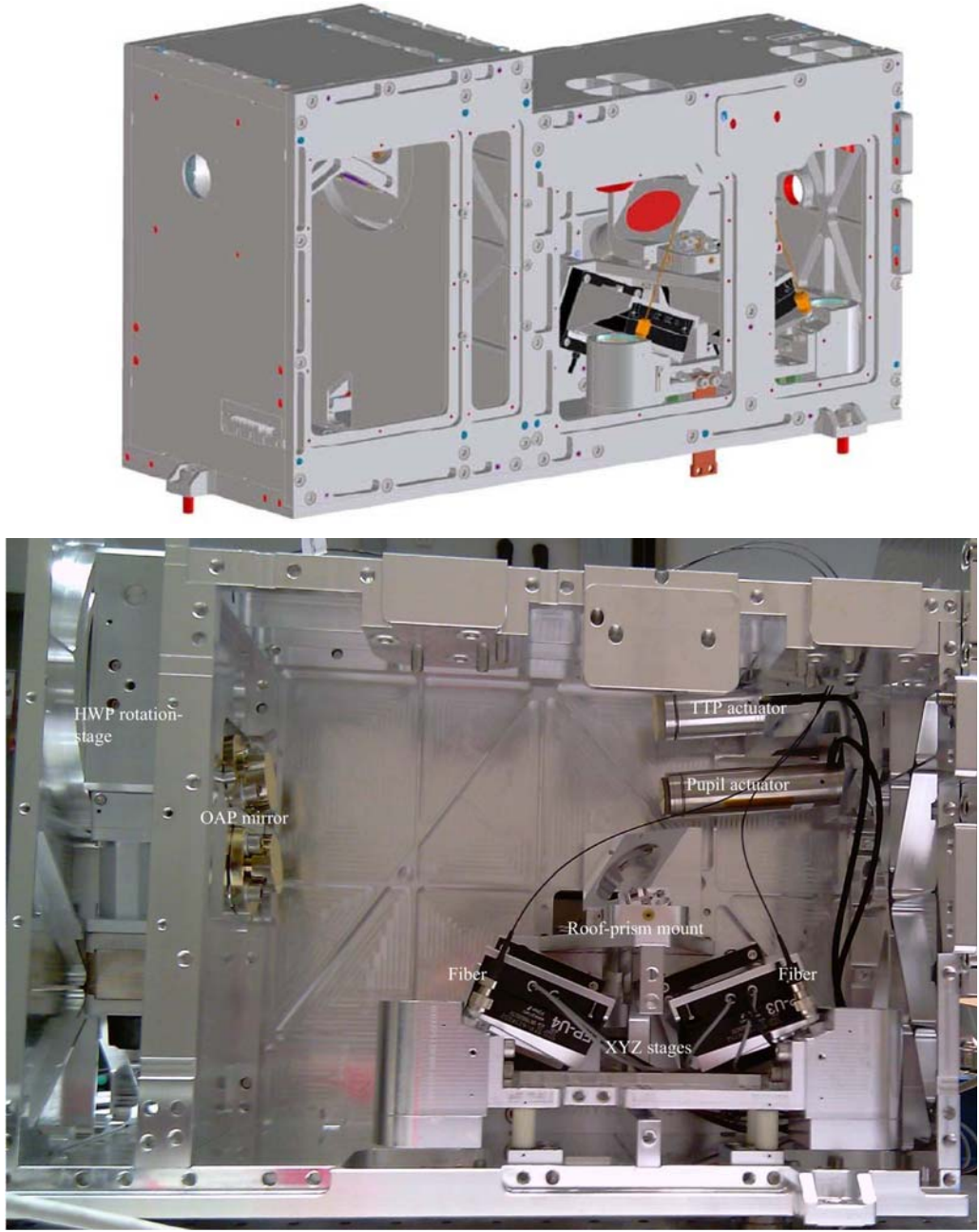

Figure 5. Top: 3D drawing of the assembled fiber coupler. The light from the telescope enters on the left. Bottom: Picture of the partially assembled fiber coupler. Not included in the picture is the K-mirror. Still missing in the picture is the dichroic and the roof-prism.

Figure 5 shows a CAD model of the final unit. The backside of the box provides the mechanical interface with the guiding system receiver. The whole structure will be screwed to the optical bench in the cryostat. Figure 5 (bottom) shows the almost fully assembled fiber coupler.

### 2.5.2 Diamond-turned mirrors

All off-axis parabolas are made from diamond-turned aluminum. Since aluminum is too soft to be post-polished, a nickel layer is deposited on the substrate. On top of the nickel, a layer of gold ensures high reflectivity in the infrared. In total, each fiber coupler contains seven off-axis parabolas. The mirrors M5 and M7 (see Figure 1) are identical off-axis parabolas with a radius of curvature of 400mm and an off-axis distance of 42mm. The mirror M9 with a curvature radius of 200mm and an off-axis distance of 40mm focuses the beam on the roof-prism. M11 is the most extreme off-axis parabola, with an off-axis distance of 105.5mm and a curvature of 200mm. The mirror M12 determines the magnification at the fiber tip, with a curvature radius of 104.35mm and an off-axis distance of 28mm. The correct alignment of the mirror is ensured by pins together with flats at the mirror body that fix the orientation. A first set of mirrors was delivered to MPE end of 2011. All mirrors turned out to be very good in terms of surface accuracy and surface roughness. Figure 6 shows one of the parabolic mirrors before mounting and the measured surface accuracy. Given the surface accuracy of $\lambda/4$ P-V at 633 nm ($\lambda/20$ RMS) and a roughness of Rq = 1.69nm over 0.3mm test-length, the mirrors provides a suitable optical quality for our purpose.

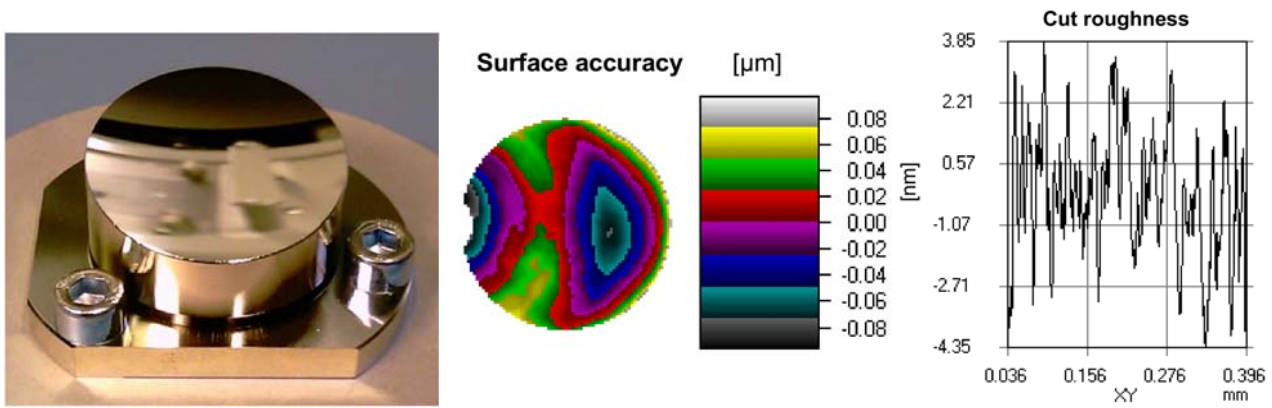

Figure 6. Picture of one diamond-turned parabolic mirror, as delivered (left). The reflective coating is hard gold without an additional protective layer. The orientation of the mirror is fixed by the flat reference surfaces at the side and alignment pins in the fiber coupler. The 2D surface accuracy (middle) is $\lambda/4$ P-V (@633nm over the full aperture. The surface roughness (right) over 0.4mm is Rq = 1.69 nm (measurements courtesy to Kugler GmbH).

### 2.5.3 Tip/tilt/piston actuator

The tip/tilt/piston actuator is a Physik Instrumente S-325.3SL piezo actuator (see Figure 5). Each unit is equipped with an internal feedback sensor. Given that the actuator operates at 240K and in vacuum, the normal capacitative sensors were replaced by strain-gauge sensors. Several units of this type have already been tested and characterized at MPE. The full travel range of the actuator is 30μm, or almost 60μm of optical path at a ~1nm resolution. In terms of tilt, the full stroke is 4mrad or 8mrad optical tilt at a 0.2μrad resolution, corresponding to a stroke of 3.7" and 0.09mas resolution on sky. Therefore the full FoV of the VLTI can be accessed, while leaving some margin for alignment corrections. The 60μm OPD stroke is sufficient to compensate fast piston perturbations. Typically the OPD varies by a few micron on the timescale of a second[6]. Larger OPD variations have to be offloaded to the main delay lines. Since the fringe-tracking and the tilt-correction loops require a fast actuator, the frequency response of the unit is an important design parameter. We have tuned the control parameters to optimize the transfer function. Using the internal strain-gauge sensors, the transfer function of each unit was measured. Figure 7 shows a typical transfer function with a 3 dB cutoff around 300Hz. The deviations among the individual units are negligible.

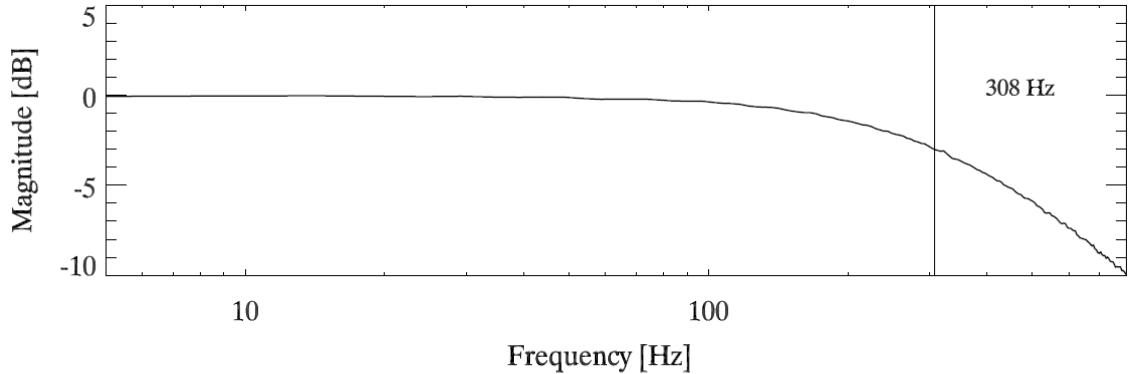

Figure 7 Transfer function of one tip/tilt/piston actuator, measured with the internal strain-gauge sensors of the unit. The 3 dB cutoff power, i.e. where the response is attenuated by a factor two, is located at 308 Hz.

**2.5.4 Pupil actuator**

The pupil actuator is a Physik Instrumente S-330.8SL piezo (see Figure 5) actuator with a 10mrad tilt stroke (20mrad optical tilt). Located at the focus of a parabola with 200mm focal length, the actuator tilt can be converted into a pupil adjustment stroke of $20 \cdot 10^{-3} \cdot 200mm = 4mm$. Given that the pupil diameter in the VLTI lab is 18mm, the relative stroke is 22%. The actuator is required to compensate the pupil wander that occurs between the telescope and the lab. The main causes of the pupil wander are the delay line wobble and the telescope run-out, which can move the pupil by ~1mm. However the pupil wander happens on timescales of minutes and therefore the actuator response is less critical.

**2.5.5 XYZ fiber positioner**

The fiber positioner is a Physik Instrumente P-625.2CL translation stage (see Figure 5) that is required to center and focus the fiber on the object. The full travel range along three axes is 0.5mm at a few nm resolution. This leaves some alignment margin, given that the nominal field diameter at the fiber position is 0.2mm (2" FoV UT). The fine resolution easily allows centering and focusing the fiber with its mode-field radius of 3.8μm.

## 3. GUIDING SUBSYSTEM

**3.1 Overview**

The purpose of the tip/tilt control-loop is to correct image jitter introduced by the VLTI tunnel atmosphere. The perturbations of the tunnel atmosphere have been studied extensively. One of the major outcomes is that the perturbations are dominated by the lowest order optical disturbance, i.e. tip/tilt, that accounts for 90% of the tunnel atmosphere variance[7]. Naturally the perturbations introduced in the tunnel depend on the optical path a beam has to travel, ranging from a few ten up to a few hundred meters. Another factor are the ambient wind conditions, that fuel the turbulence in the tunnel. A median profile of the perturbations has been established by Gitton & Puech[6]. They describe the tip/tilt disturbance as a broken powerlaw with a break frequency around 1 Hz. Figure 5.2 shows their tunnel atmosphere model. The integration of the model yields a 1-axis tip/tilt RMS of 38mas on sky (UT) during a 100 s exposure time. The corresponding 2-axis tip/tilt is about a factor sqrt(2) greater, i.e. 54mas on sky (240mas in case of the ATs).

**3.2 Tip/tilt injection losses**

The loss as function of tilt relative to the fiber mode-field radius has been published by Wallner et al.[10]. We used their loss curve and converted it to an equivalent angle on sky. In case of the UTs (D = 8m), the corresponding angle is $\omega_B/f$ = 41mas at $\lambda = 2.15\mu m$. For the ATs (D = 1.8m), the equivalent angle is 181mas. Using the mode-field equivalent angle on sky together with the loss curve of Wallner et al.[10] the average loss as function of tilt error on sky can be retrieved. Since the atmospheric jitter introduces a tilt fluctuation rather than a static offset, only the loss as function of tilt RMS is of interest. For that we calculated the average loss for a tilt distribution of 10k random samples with a certain RMS. Figure 8 shows the coupling efficiency relative to the nominal coupling for various values of the tilt RMS. Given that the tunnel atmosphere model predicts a 2-axis tilt RMS of 54mas on sky (UT), the flux loss is tremendous. Only about 50% of the nominal flux is actually coupled into the fiber because of the tip/tilt turbulence.

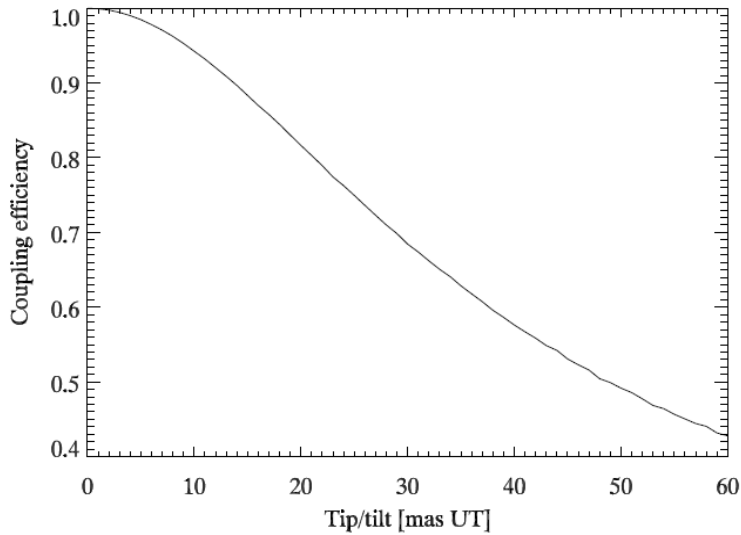

Figure 8. The loss of coupling efficiency relative to the nominal coupling is shown as function of tilt jitter RMS (UT).

### 3.3 Tip/tilt laser beacons

One tip/tilt laser beacon is launched at the STS of each telescope. The beacon, supposed to track field motion, has to be injected in a field plane. Since injecting the laser should not vignette or add any optical elements to the VLTI optical train, the injection is done by focusing the laser onto a field mirror of the STS. This creates a bright scattering spot in the respective plane. In this way, an artificial guide star is introduced in the FoV. Tracking on this artificial star therefore allows stabilizing the image motion. The laser beam is launched at a shallow angle (5.8°) from below the optical axis. Figure 9 shows the injection principle. A shallow injection angle is favored because the scattering efficiency strongly depends on the angle between the specular reflection and the scattering direction. Nearly all the scattered power is radiated away in a small cone (< 10°) around the specular reflection. In our injection scheme, the specular reflection itself will be completely baffled within the star separator enclosure.

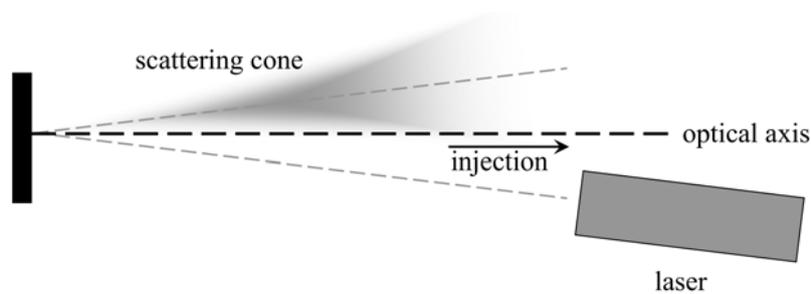

Figure 9. Sketch of the laser injection. The laser is launched below the optical axis and focused onto the field mirror. Only the light scattered in the direction of the optical axis is injected in the VLTI train. The specular reflection is baffled inside the STS housing.

The optical design of the tip/tilt laser beacon is such that it contains only off-the shelf optical components. The laser source is a commercial fiber-pigtailed laser diode with a wavelength of 658 nm and a nominal output power of 60mW. A constant current laser driver housed in the STS LCU provides the power for the laser. The laser can be switched on and off, via a digital I/O. The laser fiber is connected via a FC connector to a beam collimator, creating an output beam of 7mm diameter. The collimator is mounted in a lens tube system. Mounted in an adjustable extension of the lens tube, a f = 600mm lens focuses the beam on the field mirror of the STS. The length of the extension can be adjusted by an internal thread to focus the beam. The whole lens tube system sits on a kinematic mount, that can be tilted along two axes. The mount allows an adjustment of the laser pointing by ±4° at a resolution of ~3'. This allows centering the scattering spot on the mirror. If the beam is centered, the nominal launch angle is 5.8° with respect to the optical axis. The whole assembly is mounted on a magnetic plate that can be removed from the STS without loosing the initial alignment. Figure 10 shows one tip/tilt launcher installed on a UT STS. The laser points to the corresponding field mirror.

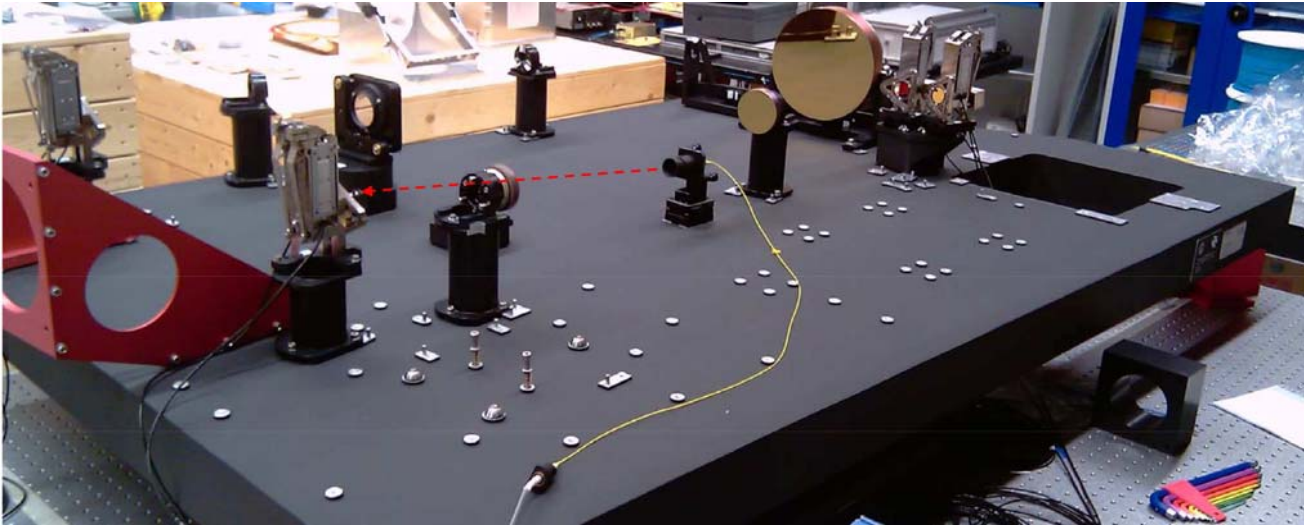

Figure 10. Picture of one tip/tilt laser launcher installed on a UT STS. The launch optics is fed by a fiber-pigtailed laser diode (yellow). The red arrow indicates the pointing direction of the beam that is focused on the field mirror.

### 3.4 Guiding receiver

#### 3.4.1 Optical design

The tip/tilt guiding laser injected in the FoV travels the same optical path through the VLTI and the beam combiner instrument as the stellar light. As described in Chapter 4, the beam propagates through the fiber coupler up to the H/K dichroic. While the science wave band is reflected to the fiber injection optics, the laser wavelength passes the dichroic and exits the fiber coupler in an 18mm collimated beam. The guiding receiver, which is mounted like a backpack on the fiber coupler, picks up the laser beam and images the beacon on a position sensitive diode (PSD). The beam pickup happens with another dichroic, which reflects the tip/tilt guiding laser and lets the pupil guiding laser and the stellar H-band light pass to the guiding receiver. We chose a silicon substrate dichroic since silicon is opaque at wavelengths shorter than 1100nm. Together with a reflective coating it provides a high reflectivity (99%) of the 658nm laser line and ensures that no stray-light is propagated to the acquisition camera. The pupil laser (1200nm) and the stellar acquisition band (1.5 − 1.8μm) are able to pass the dichroic with negligible reflection losses (0.5%). Figure 11 shows the optical design of the receiver. The reflected laser light is focused by an f = 400mm lens onto the PSD. Given the focal length, the $4 \times 4mm^2$ active area of the diode corresponds to 0.573° or an equivalent FoV (UT) of ±2.3" on sky. Thus the FoV of the guiding receiver is a factor two larger than the nominal VLTI FoV. Therefore enough stroke margin is available to ensure that the artificial laser beacon is always visible on the guiding receiver.

#### 3.4.2 Mechanical design

The front of the guiding receiver structure is screwed to the fiber coupler unit. The structure is completely manufactured from aluminum, i.e. the same material as the fiber coupler. This avoids differential contraction of the material, when the unit is cooled down to the operating temperature of 240K. Cylindrical pins ensure the correct alignment of the receiver and the fiber coupler. Figure 12 shows the assembly during testing. The large size of the box results from the long focal length of the lens, which provides a useful FoV on the diode. All optics, with the exception of the dichroic are off-the-shelf products. The fused silica lens is centered by three springs oriented at 120° with respect to each other. The springs are directly cut out of the mounting plate. The two fold mirrors provide the long focal distance. One of the fold mirrors sits on a tilt stage that allows aligning the beam with the diode. The sensor, a position sensitive diode, is mounted on a aluminum plate, that can be rotated by a few degrees. This allows aligning the diode coordinate system with the fiber coupler coordinate system that is defined by the orientation of the tip/tilt actuator.

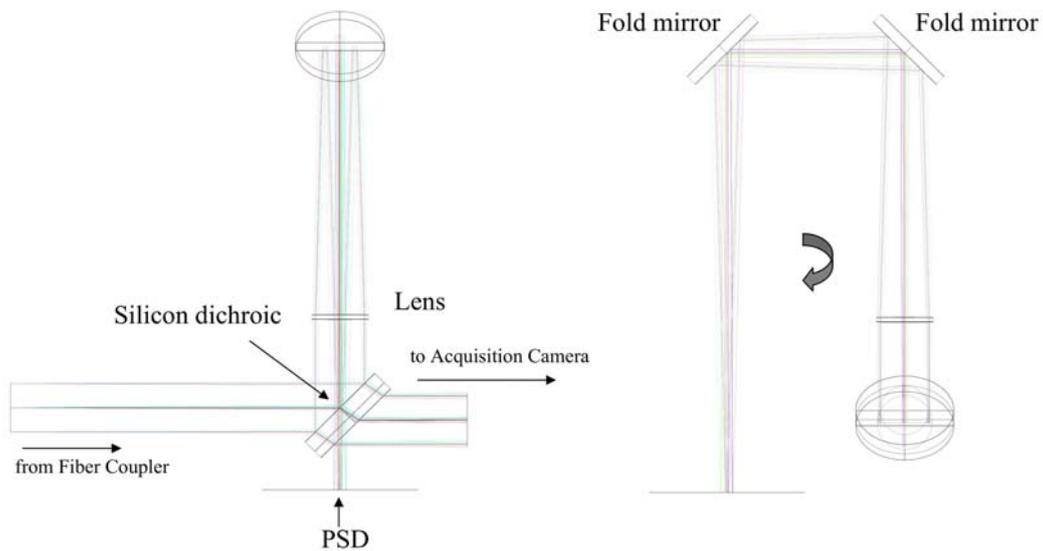

Figure 11. Optical design of the guiding receiver. The collimated beam exiting the fiber coupler (see Figure 1) is split by a silicon dichroic in the guiding receiver. The 658 nm laser light is reflected, relayed via flat mirrors and focused with a commercial f = 400mm lens on the position sensitive diode. The pupil guiding laser (1200 nm) and the stellar H-band light pass the dichroic to the acquisition camera. Since silicon is opaque at wavelengths shorter than 1100 nm, no laser stray-light is able to contaminate the propagated beam.

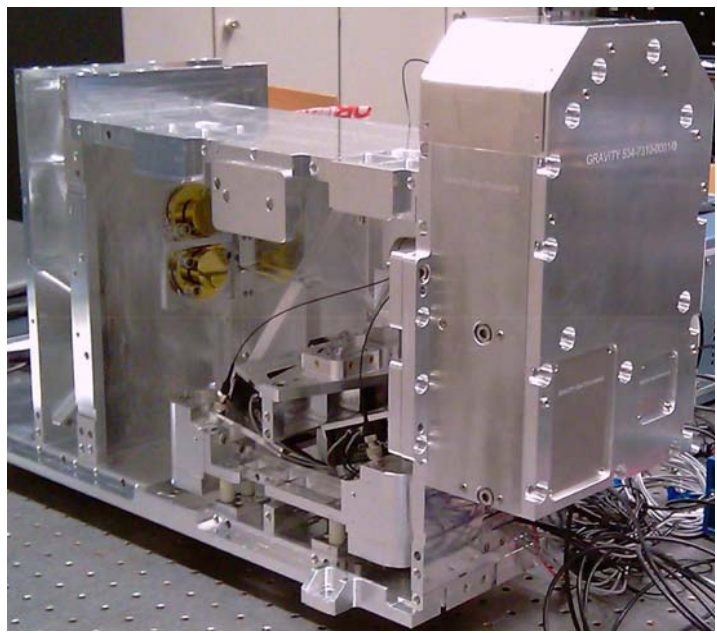

Figure 12. Guiding receiver mounted on the fiber coupler. In this picture, the beam exit of the guiding receiver is closed with a small aluminum plate. The connector cable for the diode enters the unit from the bottom.

### 3.4.3 Sensor

The tip/tilt guiding beacon is tracked with a position sensitive diode (PSD). A PSD consists of a uniform resistive layer formed at one or both surfaces of a high-resistivity semiconductor substrate, and a set of electrodes formed on the ends of the resistive layer for extracting position signals. The active area, which is also a resistive layer, has a PN junction that generates a photo-current if light hits the surface. The photo-current that is measured at a particular electrode is inverse proportional to the distance between the light spot and the electrode due to the increasing resistance. By comparing the photo-currents at the output electrodes, the spot position can be derived. Figure 13 illustrates the position measurement. The photo-currents at the electrodes are amplified and measured as output voltages. The PSD used in the guiding receiver is a silicon duo-lateral PSD from Pacific Silicon Sensors. Its four output voltages are amplified and converted into X-Y position voltages by an amplifier board developed by MPE. The amplifier is optimized for low-noise performance at very high amplification with a 3 dB bandwidth of 820Hz. An on-board sample-and-hold chip allows storing and subtracting the dark current of the diode. This is necessary since the diode is operated at very low light levels. In this regime, leakage currents in the PSD are not negligible and have to be subtracted. The calibrated signals are then converted into an X and Y voltage via the analogue circuitry.

Taking into account the scattering efficiency, the VLTI transmission and the GRAVITY instrument transmission, only about 1μW of the initial 60mW laser reach the PSD. However, at that level, the position resolution of the PSD is limited by material inhomogeneities to about $dL/L = 0.1\%$. This can be converted into a position resolution on sky by multiplication with the equivalent FoV of 2.3", resulting in a sensor noise of ~2.3mas RMS (UT).

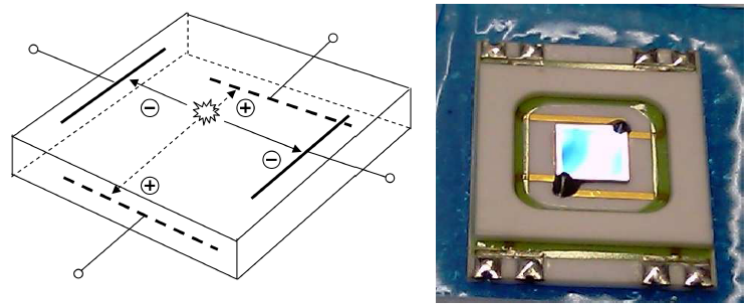

Figure 13. Illustration of the working principle of a position sensitive diode (Hamamatsu, 2012). The incident light spot generates a photo-current. The resistance between the electrodes and the spot position is proportional to the distance. Therefore by comparing the currents of two electrodes derives the spot position along one axis. By measuring the electron and the hole currents in perpendicular directions, the 2D position can be retrieved. Right: Image of the silicon PSD used in the guiding receiver. The active area is $4 \times 4mm^2$.

### 3.5 Control-loop performance

The tip/tilt guiding system consists of two sensors, PSD and acquisition camera[4], commanding a common actuator. Its purpose is to correct the tip/tilt jitter induced by the tunnel atmosphere and to optimize the light injection into the single-mode fibers. The actuator is a piezo-driven mirror, installed in the fiber coupler (see Sec. 2.5.3). The acquisition camera images the FoV on sky and provides a reference for the fiber position at a frame rate of ~1Hz. The low frame rate is set by readout of the detector and the object faintness. Thus the camera is only able to monitor slow tilt drifts. High frequency perturbations are therefore sensed by the PSD described in Section 3.4.3. The device locks on the artificial laser star injected at the star separator (Section 3.3). With the amplifier tuned to a bandwidth of 820 Hz, it is able to measure and correct high frequency image jitter. However, the PSD sensor is intrinsically prone to drift on slow timescales. This can either happen due to electrical drifts of the amplifier and the op-amps therein or due to a drift of the laser spot on the star separator. For this reason, a high-pass filter is used to block frequencies below 0.05Hz. In this scheme, the control-loop senses the high frequency jitter with the PSD but ignores the potential drifts of the device. The jitter stabilized image of the acquisition camera is then used to center the science and fringe-tracking object on the fiber. The fast control-loop, namely the PSD amplifier, filter and PD controller are implemented in analogue electronics. The obvious advantage is that this saves software and hardware implementation of the fast PSD sensor in an ESO LCU based real-time system. The slow acquisition camera will be operated on the instrument control workstation since at the given frame rate it does not need real-time functionality. During operation each sensor can be independently switched on or off. It is also possible to use the actuator in blind mode with both sensors switched off. This makes the implementation and testing of the system easy and adds flexibility. The acquisition camera setpoint (i.e. the star position on sky) and the blind offsets can be set via the ICS. Figure 14 shows the control-loop scheme.

The open-loop transfer function of the combined acquisition camera and the PSD system can be written as:

$$G_{sys,ol}(s) = \left( \frac{1}{(1+\tau_{PSD})^2} \cdot \frac{\tau_{fi}s}{\tau_{fi}s+1} \cdot G_{PD}(s) + \frac{1-e^{-Ts}}{Ts} \cdot \frac{1-e^{-Ts}}{s} \cdot e^{-\tau s} \cdot G_{PI}(s) \right) G_{act}$$

The camera response is of discrete nature. However it is possible to approximate the response in a continuous representation using an average signal during the camera integration time $T$, a sample-and-hold term and an additional short delay $\tau$ introduced due to computing overheads (see Hardy[8]). The PSD response can be modeled as a low-pass filter, where the amplifier bandwidth is realized by $f_b = 1/(2\pi \tau_{PSD})$. The high-pass filter with the lower frequency cutoff $f_c = 1/(2\pi \tau_{fi})$, the PD controller response $G_{PD}(s)$ and the PI controller response $G_{PI}(s)$ are standard textbook examples taken from e.g. Franklin et al.[9]. The actuator response $G_{act}(s)$ is simply the measured transfer function (see Figure 7).

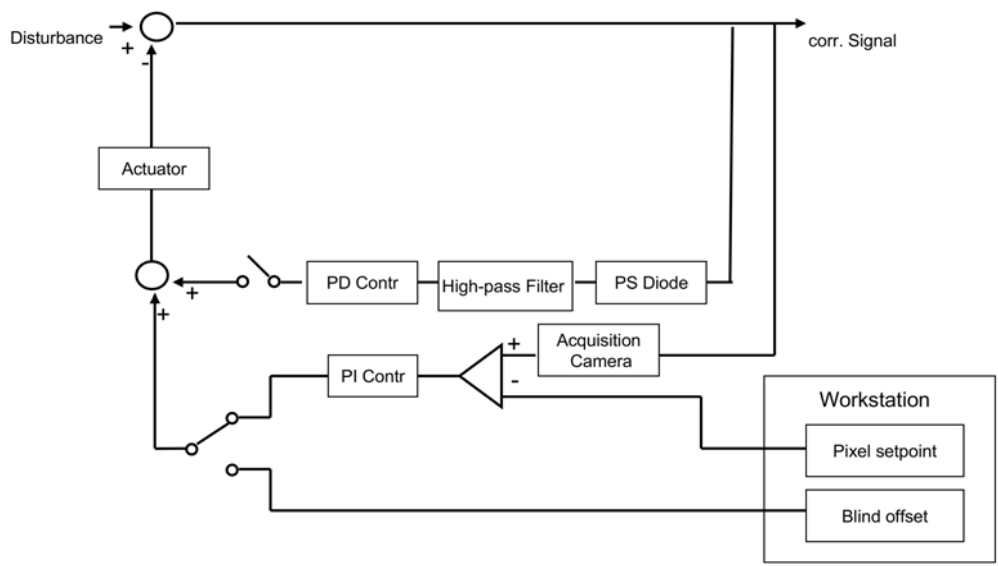

Figure 14. Control-loop scheme of the tip/tilt guiding system. Two sensors, the PSD and the acquisition camera covering high frequency and low frequency perturbations, command a common actuator in the fiber coupler. The PSD chain is implemented in analogue electronics, including the amplifier and the PD controller. An analogue high-pass filter effectively blinds the PSD for low frequency drifts, while the slow frame rate of the acquisition camera renders it insensitive to fast jitter. Both control-loop branches can be independently switched on or off. By sending a command via ICS it is possible to change the setpoint of the acquisition camera or to send a blind offset to the actuator.

The following input parameters were used for the model:
- acquisition camera integration time T = 1s
- computational delay $\tau$ = 0.5s
- bandwidth of PSD amplifier $f_b$ = 820Hz
- high-pass filter cutoff $f_c$ = 0.05Hz
- acquisition camera sensor noise (open loop) $N_{acq}$ = 0.5mas RMS
- PSD sensor noise (open loop) $N_{PSD}$ = 2.3mas RMS

The integration time was based on the fastest possible full frame read of a HAWAII detector, while the computational delay was estimated. Given the little computational effort consisting of reading a 4Mpixel frame and determining the image shift by fitting a point source, the delay seems achievable. In fact, the simulation showed that a delay shorter than 0.5 s does not change the system performance significantly. However at delays of 1s and more the performance degrades notably. Therefore in the future software implementation, the delay length has to be considered. During the fine-tuning of the model it became clear, that the high-pass filter cutoff had to be set to ~0.05Hz. The optimum cutoff is directly related to the frame rate of the acquisition camera. In simple terms, the cutoff timescale (20s) has to be significantly longer than the camera integration time (1s). If the timescale is not at least a factor 10 greater than the frame rate, the system actually becomes instable for non-zero gains. Using the input parameters in our model lead to the closed-loop performance shown in Figure 15. The modeled system attenuates disturbances between 0.1 Hz and 200Hz by at least 15 dB. Only around 0.05Hz, i.e. the filter cutoff, the attenuation is not as good. In this regime, the correction bands of the acquisition camera and the PSD overlap, leading to a degraded performance.

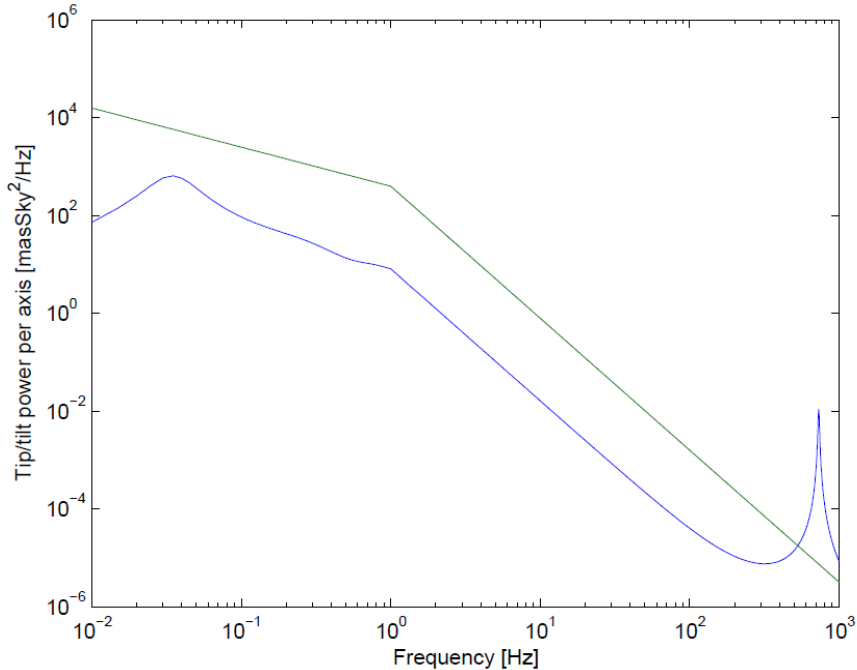

Figure 15. Power-spectrum of 1-axis tip/tilt perturbation of the VLTI tunnel without correction (green) and after correction by the guiding system (blue). The integrated disturbance of 38mas RMS before correction is reduced to 7.0mas RMS after correction.

Applying the system response to the tip/tilt power-spectrum yields the residual tip/tilt after closed-loop correction. The corresponding spectrum is shown in Figure 15. Integrating the corrected power-spectrum yields a residual tip/tilt of 7.0mas RMS. Taking into account the sensor noise, the 1-axis residual tip/tilt increases to a total of 7.5mas RMS. Compared to the 38mas RMS (2-axis: 54mas) without correction, the guiding system is able to reduce the tip/tilt perturbation in the tunnel by a factor five. According to the GRAVITY AO simulations performed by Clenet & Gendron[11], about 10mas RMS tip/tilt are not corrected by the AO. Thus the guiding system reduces the tunnel tip/tilt to a level, that the AO residuals are limiting the injection efficiency. The quadratic sum of the two contributions is about 12.5mas RMS in total (2-axis: 17.7mas RMS). As described in Section 3.2, the uncorrected tunnel tip/tilt leads to a reduction of the injection efficiency by 50%. Comparing the 2-axis residual tip/tilt with the loss curve in Section 3.2, implies that the injection efficiency can be increased to about 85% if the guiding system is used to correct the tunnel atmosphere. The remaining 15% losses originate mainly from the uncorrected AO tip/tilt.

# 4. CONCLUSIONS

The fiber coupler concept and opto-mechanical design developed in the course of this thesis ensures an efficient single-mode coupling of the telescope beams into the beam combiner instrument. Each fiber coupler unit provides the necessary functions to rotate the field on sky, to adjust the linear polarization orientation, to fringe-track, and to adjust the field and the pupil positions of one telescope beam. It splits the beam with a dichroic and feeds the science band $1.95 - 2.45\mu m$ into single-mode fibers, while the acquisition band $1.5 - 1.8\mu m$ as well as the guiding lasers are propagated to the acquisition camera and the guiding receiver. A special roof-prism allows splitting the field on sky for dual-field interferometry or partitioning the beam for single-field interferometry. The design ensures a diffraction limited beam and an optimum throughput. At the time of writing the first fiber coupler unit is almost completely assembled. Most of the optics and especially the critical optical components, such as the roof-prism, the off-axis parabolic mirrors and the half-wave plate have already been delivered to MPE. The H/K dichroic is not yet available but a contract has been placed. The main results of the fiber coupler development are:

- The fiber coupler optical design is close to aberration free. It features a nominal K-band Strehl ratio of > 99% across the 2" FoV of the VLTI. This exquisite image quality is achieved by a combination of off-axis parabolic mirrors to avoid field aberrations and the absence of lenses to avoid chromatic aberrations.

- A corner stone of the optics is the roof-prism. Although only $1.5 \times 4mm^2$ in size, the prism layout allows to either split the FoV in the dual-field operation of GRAVITY or to use it as a beam splitter in the single-field operation.

- The optical design based on aluminum off-axis mirrors has many advantages compared to a lens-based design, such as reduced costs, a small overall volume and the absence of chromatic aberrations. Since the off-axis mirrors and the fiber coupler structure are made of aluminum, the design is also temperature invariant and requires no re-focusing once cooled down. Poor surface accuracy and roughness prevented the widespread application of diamond-turned aluminum mirrors in optical systems until a few years ago. However, the off-axis mirrors supplied to MPE with a surface accuracy of $< \lambda/20$ RMS at 633nm and a roughness of $R_q = 1.7$ nm are suitable for near infrared optics.

The tip/tilt guiding system uses an artificial laser beacon to correct high frequency image jitter occurring in the VLTI tunnel. Low frequency image drifts are sensed on sky by the acquisition camera. The combination of a laser beacon and on sky guiding provides an efficient way to correct the tip/tilt power-spectrum in the tunnel and to optimize the coupling efficiency into the single-mode waveguides of the beam combiner instrument.

- The high frequency tip/tilt perturbations in the VLTI tunnel can be corrected with an analogue position sensitive diode tracking on an artificial laser beacon. The beacon itself is introduced by using the scattered light of a 658 nm laser focused onto a fieldplane of the star separator.

- The combination of the acquisition camera sensing slow image drifts and the position sensitive diode sensing high frequency jitter allows correcting most of the tunnel atmosphere. The system reduces the typical 54mas 2-axis RMS (UT) tip/tilt to a residual 7.5mas RMS. The specification of < 15mas is clearly met with the proposed design.

- The compensation of the tunnel tip/tilt significantly increases the coupling efficiency of the single-mode fibers. In the current state, i.e. without beam stabilization, on average only about 50% of the light is coupled into the fibers. The tip/tilt guiding system can increase the effective coupling to 85%. The remaining 15% are lost due to AO tilt residuals, which can not be recovered with the guiding system.